\documentclass[aps,prb,reprint]{revtex4-1}

\usepackage{graphicx,hyperref,amsmath,epstopdf,natbib,textcomp,color,braket}

\begin{document}


\title{Addressing spin transitions on $^{209}$Bi donors in silicon using circularly-polarized microwaves}



\author{T. Yasukawa}
\affiliation{Department of Electrical Engineering, Princeton University, Princeton, New Jersey 08544, USA}

\author{A. J. Sigillito}
\email[]{asigilli@princeton.edu}
\affiliation{Department of Electrical Engineering, Princeton University, Princeton, New Jersey 08544, USA}

\author{B. C. Rose}
\affiliation{Department of Electrical Engineering, Princeton University, Princeton, New Jersey 08544, USA}

\author{A. M. Tyryshkin}
\affiliation{Department of Electrical Engineering, Princeton University, Princeton, New Jersey 08544, USA}

\author{S. A. Lyon}
\email[]{lyon@princeton.edu}
\affiliation{Department of Electrical Engineering, Princeton University, Princeton, New Jersey 08544, USA}


\date{\today}

\begin{abstract}
Over the past decade donor spin qubits in isotopically enriched $^{28}$Si have been intensely studied due to their exceptionally long coherence times. More recently bismuth donor electron spins have become popular because Bi has a large nuclear spin which gives rise to clock transitions (first-order insensitive to magnetic field noise). At every clock transition there are two nearly degenerate transitions between four distinct states which can be used as a pair of qubits. Here it is experimentally demonstrated that these transitions are excited by microwaves of opposite helicity such that they can be selectively driven by varying microwave polarization. This work uses a combination of a superconducting coplanar waveguide (CPW) microresonator and a dielectric resonator to flexibly generate arbitrary elliptical polarizations while retaining the high sensitivity of the CPW.

\end{abstract}

\pacs{}

\maketitle


Donors spins in Si are among the most promising quantum bits owing to their long coherence times ($T_{2}$) which exceed seconds in isotopically enriched $^{28}$Si \cite{tyryshkin2012, Muhonen2014, Saeedi2013, Steger2012}. Bismuth donor electrons are particularly attractive because they have clock transitions which are first order insensitive to magnetic field noise \cite{wolfowicz2013, mohammady2010, mohammady2012, george2010, mortemousque2014}. This means that even in natural Si, electron spins can have long coherence times \cite{wolfowicz2013}. In Bi doped Si clock transitions come in pairs of nearly degenerate transitions separated by $\sim$1 MHz. These are predicted to be excited by microwaves of opposite circular polarization \cite{mohammady2012, pica2015surface}. In this work we combine a coplanar waveguide microresonator (CPW) with a dielectric resonator to generate microwaves with tunable polarization. By varying this polarization, we demonstrate the selective addressability of the 7.03 GHz clock transitions. This will be important for hybrid donor-dot quantum computing schemes since the 5 GHz clock transition was recently predicted to form an avoided crossing with silicon based quantum dots\cite{pica2015surface}. This was discussed in the donor-dot surface code proposal by Pica et \textit{al}\cite{pica2015surface} which also suggested addressing the clock transitions using microwave polarization.

Bismuth donors in Si have a large hyperfine coupling which not only gives rise to a large zero field splitting, but also allows for rapid manipulation of both the electronic and nuclear spin states when operating in the intermediate field regime where the hyperfine coupling is comparable to the Zeeman splitting\cite{feher1, Morley2013, bienfaitSC, george2010}. In this regime we describe the states using the total spin, $F$, and its projection, $m_F$ \cite{mohammady2010}. The total spin is given by $F=I\pm S$ where $I$ is the nuclear spin and $S$ is the electron spin. At the 7.03 GHz clock transition the two nearly degenerate transitions are described in the $\ket{F,m_F}$ basis by $\ket{5,-1}\Leftrightarrow \ket{4,-2}$ and $\ket{5,-2}\Leftrightarrow \ket{4,-1}$. In the high field limit the second transition is forbidden since it involves a nuclear spin flip, but due to strong mixing, both transitions are accessible near the clock transition. By convention we will still refer to the $\ket{5,-1}\Leftrightarrow \ket{4,-2}$ transition as allowed and the $\ket{5,-2}\Leftrightarrow \ket{4,-1}$ transition as forbidden. These transitions were discussed by Mohammady \textit{et al.} who first pointed out their addressability based on microwave polarization\cite{mohammady2012}. Because these two transitions are between four distinct states, they can be used to form a two-qubit system.

Quantum computing implementations based on donors in Si usually involve only one electron spin resonance (ESR) transition. Using an additional, nondegenerate transition would require rapid sweeping of the external magnetic field ($\vec{B}_{0}$) which is unrealistic. Alternatively one can use nearly degenerate transitions like those near the 7.03 GHz clock transition (0.6 MHz splitting). To selectively address nearby transitions, one would conventionally use slow control pulses so that the pulse bandwidth is narrow relative to the separation of the transitions. In the case of bismuth, the transitions have opposite gyromagnetic ratios ($g$) and are therefore excited by opposite microwave polarizations\cite{Schweiger2001}. By taking advantage of polarized microwaves, we can overcome the bandwidth limitations on the pulses to rapidly and selectively address the clock transitions. This technique is applicable not only to Bi donor electrons, but also nitrogen-vacancy centers\cite{alegre2007, mrozek2015} and any other system with nearly degenerate transitions having opposite $g$.

\begin{figure}[h]

\includegraphics{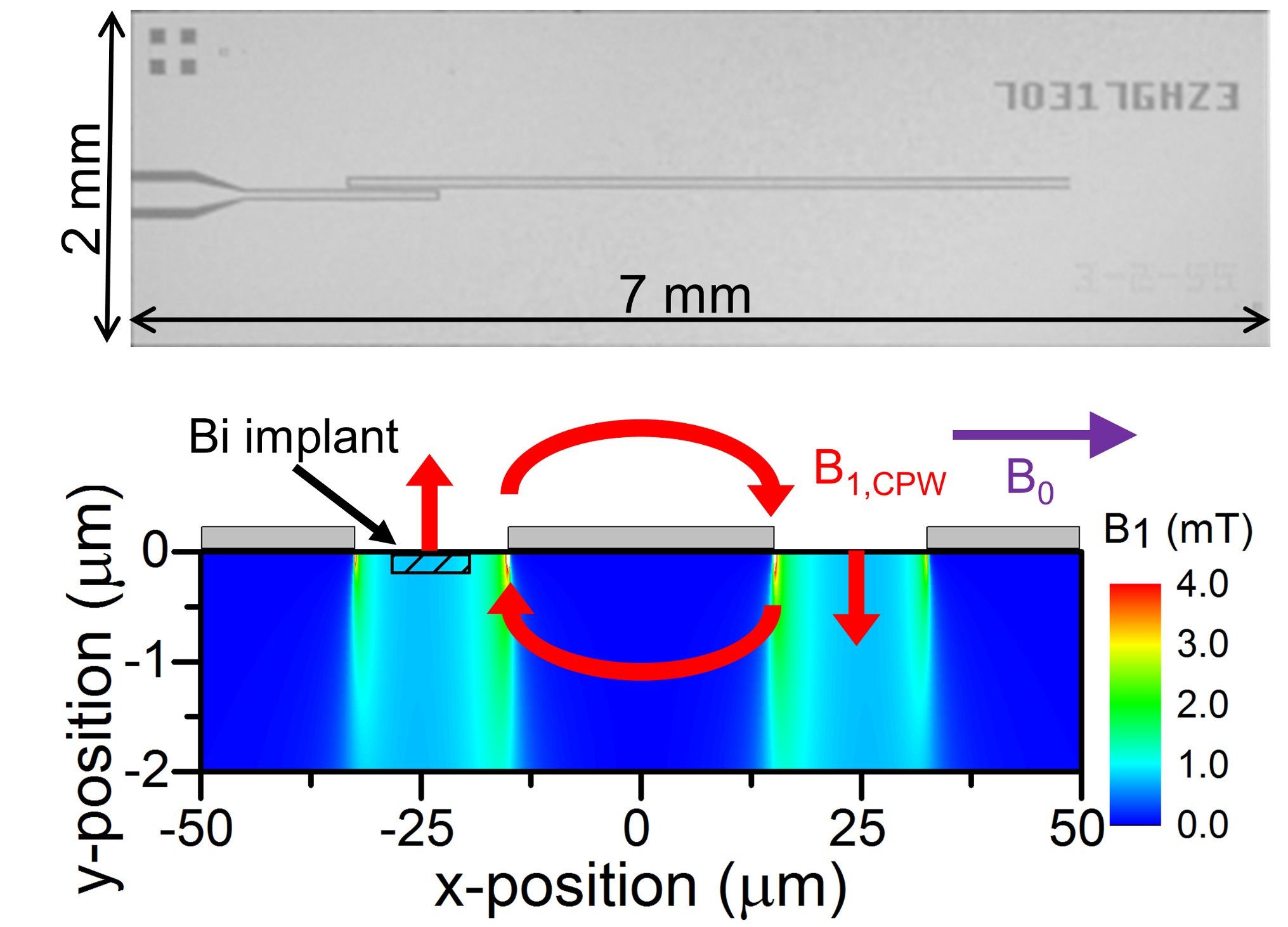}
\caption{(a) Optical micrograph of the $\lambda / 4$ shorted CPW microresonator. The center pin is 30 $\mu$m and the gap width is 17.4 $\mu$m. (b) Cross sectional density plot of the microwave magnetic field at an antinode in $\vec{B_{1,CPW}}$. Note that the two axes have been plotted with different scales. The Nb conductors are illustrated by the cartoon rectangles at the top of the plot. The $^{209}$Bi implanted region is shown as a hatched box in the gap on the left. The direction of $B_{1,CPW}$ is normal to the surface at the donors as illustrated by the red arrows.}
\label{fig:fig1}
\end{figure}

Many techniques for generating circularly polarized microwaves \cite{huchison1960, chang1964, eshbach1952} exist, but none were well suited to our application because of our sample geometry. The sample consists of a 2 $\mu$m epitaxial layer of $^{28}$Si grown on high resistivity p-type Si. The epi-layer was doped with 5$\times$10$^{15}$ P donors/cm$^{3}$ and Bi donors were implanted with a box profile to a depth of approximately 100 nm with a peak density of $10^{17}$ Bi donors/cm$^{3}$ as described by Weis \textit{et al.} \cite{weis2012}. For this small sample volume, 3-dimensional resonators will have a very small fill factor and a correspondingly small signal. Planar microresonators are particularly well suited to these kinds of samples where the spins are located near a surface\cite{malissa2012, sigillito2014}, but previously developed techniques for generating circularly polarized microwaves using planar resonators\cite{alegre2007res, henderson2008, mrozek2015} are incompatible with superconductors since they would require $B_{0}$ normal to the superconducting film. The use of normal metal would limit the Q value and thus degrade the sensitivity of the resonator. To overcome these problems, we have developed a new technique for generating circularly polarized microwaves which combines a tunable double-stacked dielectric resonator\cite{Jaworski1997, colton2009} with a superconducting coplanar waveguide (CPW) microresonator\cite{malissa2012, sigillito2014} (shown in Fig.~\ref{fig:fig1}(a)). The two resonators were arranged to have orthogonal modes (denoted $\vec{B}_{1,D}$ and $\vec{B}_{1,CPW}$ for the dielectric and CPW resonators, respectively) so that by tuning the relative phase and amplitude of the two modes, the superposition of the fields can produce microwaves of any arbitrary polarization. 

While the dielectric resonator produces a very homogeneous microwave magnetic field, CPW microresonators are notorious for their inhomogeneities \cite{malissa2012, sigillito2014} which we plot in Fig.~\ref{fig:fig1}(b)). The inhomogeneity of $\vec{B}_{1,CPW}$ will lead to a distribution in microwave polarization. To tighten this distribution the Bi donors were selectively implanted in 9 $\mu$m $\times$ 1.6 mm strips. Microresonators were then patterned directly on the sample surface as previously described\cite{malissa2012, sigillito2014} and were aligned so that the donors are centered in only one of the CPW gaps (schematically illustrated by the hatched region in Fig.\ref{fig:fig1}(b)). It is important to only dope one side of the microresonator since there is a 180$^{\circ}$ phase shift between $\vec{B}_{1,CPW}$ in the two gaps \cite{simons2001} (giving them opposite circular polarizations).

\begin{figure}[h]

\includegraphics{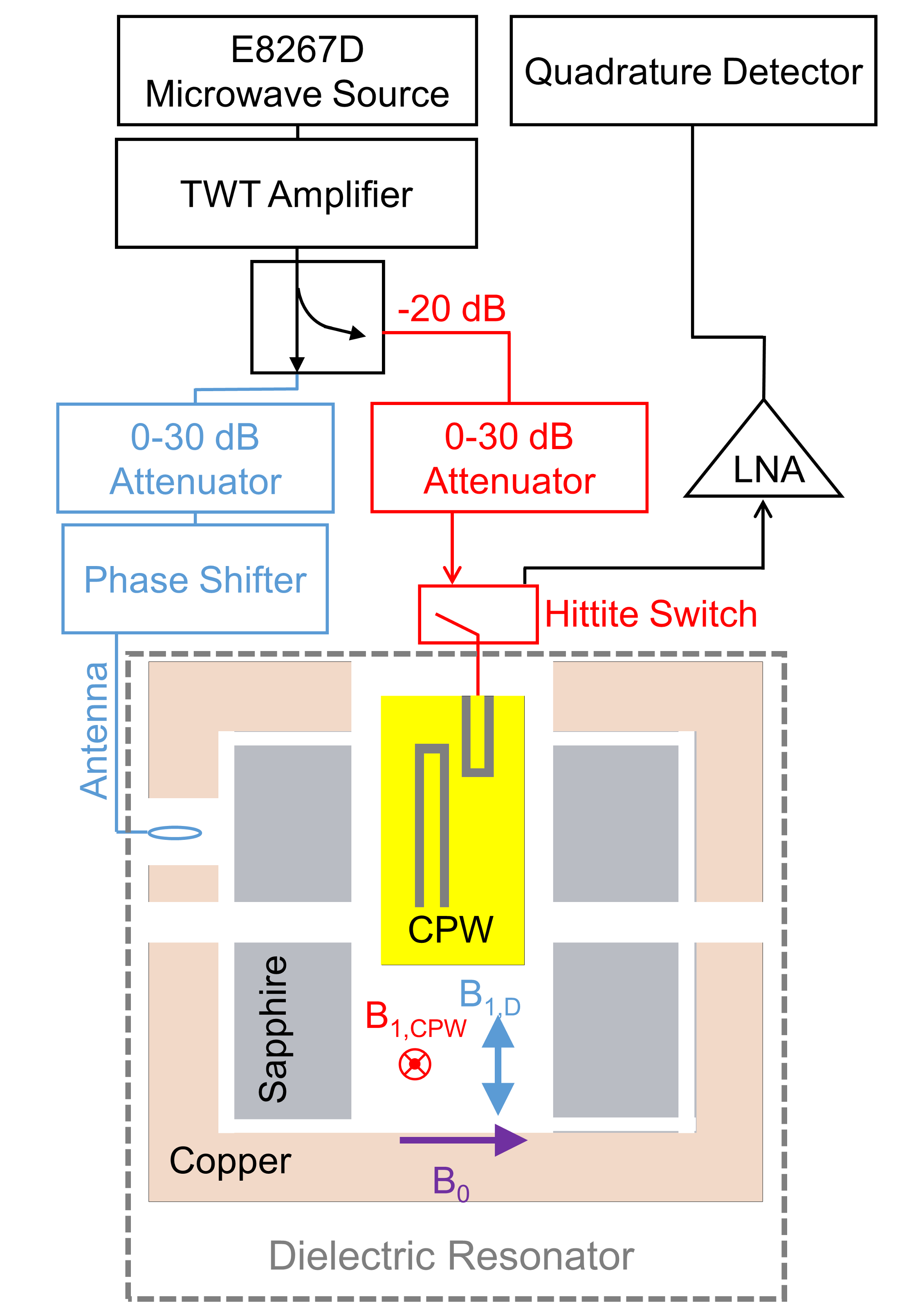}
\caption{Schematic of experimental setup including a cartoon cross section of the dielectric resonator with CPW microresonator (yellow) mounted inside. Microwaves feed into the dielectric resonator through the antenna connected to the blue arm whereas microwaves excite the CPW through the red arm. The directions of the magnetic fields at the donors are shown by the arrows. $\vec{B}_{1,CPW}$ (red) is directed into and out of the page, whereas $\vec{B}_{1,D}$ (blue) is vertical. $\vec{B}_0$ (purple) is orthogonal to both microwave magnetic fields (horizontal). }
\label{fig:fig2}
\end{figure}

The CPW microresonator was mounted coaxially inside the volume resonator as show in Fig.~\ref{fig:fig2}. The magnetic field, $\vec{B}_{0}$, was oriented in the plane of the Nb and perpendicular to the center pin. In this orientation, the component of $\vec{B_{1}}$ capable of driving spin rotations is primarily in the microresonator gap\cite{sigillito2015} and normal to the surface as shown in Fig.\ref{fig:fig1}(b). The dielectric resonator (illustrated in Fig.~\ref{fig:fig2}) has $\vec{B}_{1,D}$ oriented along the CPW center pin such that $\vec{B}_{0} \perp \vec{B}_{1,CPW} \perp \vec{B}_{1,D}$ (illustrated by the arrows in Fig.~\ref{fig:fig2}). By tuning the relative amplitude and phases of the two linearly polarized microwave fields, we can generate microwaves with arbitrary polarizations normal to $\vec{B}_{0}$.

To reliably control the microwave polarization over the duration of the experiment, it is important to minimize phase drifts. For this reason, we used a homodyne excitation scheme which is diagrammed in Fig.~\ref{fig:fig2}. A single microwave source (Agilent E8267D) was amplified using a travelling wave tube (TWT) amplifier and then split using a 20 dB directional coupler into two arms; a CPW arm (red in Fig.~\ref{fig:fig2}) and a dielectric resonator arm (blue in Fig.~\ref{fig:fig2}). Both arms were equipped with variable attenuators to control their microwave amplitudes and the dielectric resonator arm was also equipped with a phase shifter. The spin echo was detected using a cryogenic low noise amplifier connected to a quadrature detector. A Hittite switch (HMC347LP3) was used to protect the low noise amplifier from the microwave pulses.

\begin{figure}[h]
\includegraphics{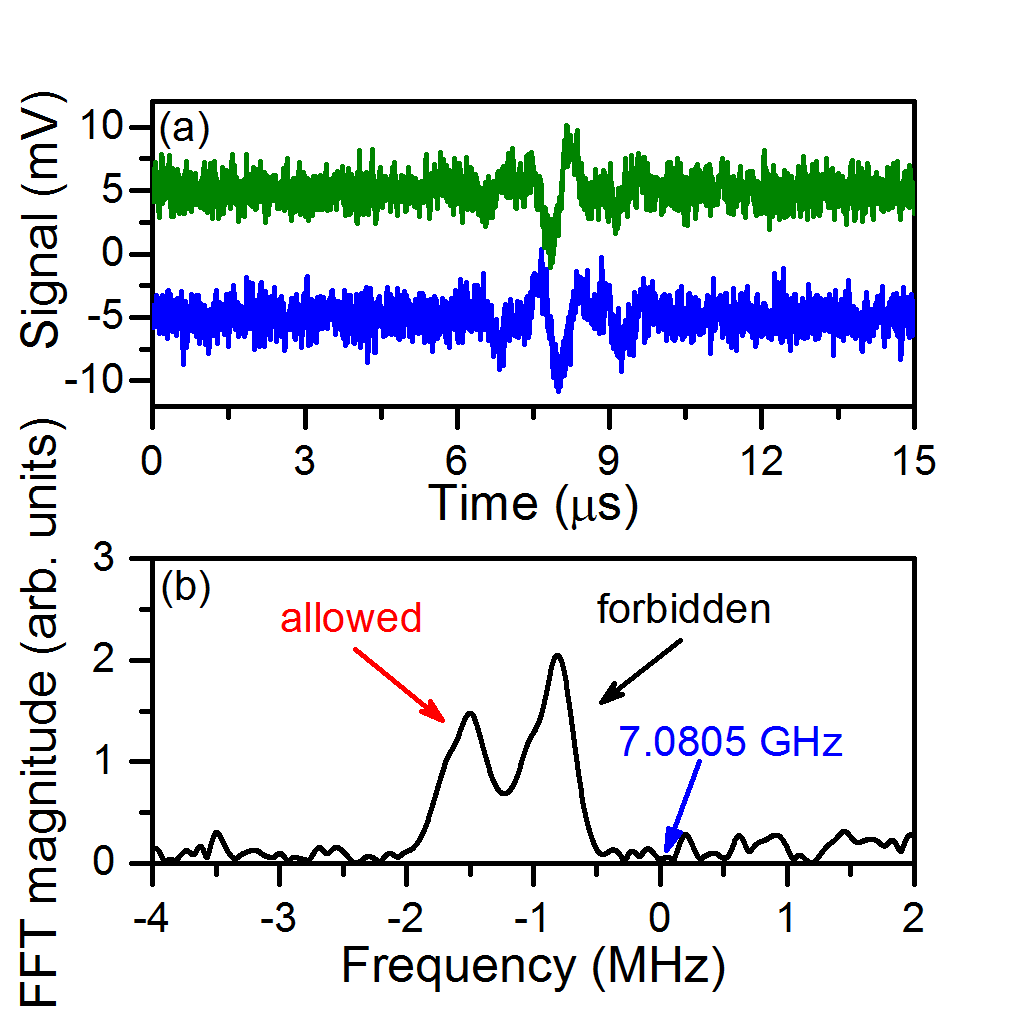}
 \caption{(a)Typical spin echo signal showing in-phase (top green) and quadrature (bottom blue) signal components. The curves have been offset for clarity. (b) Fourier transformed spectrum of the echo shape. The allowed (left) and forbidden (right) transitions are both clearly resolved. The frequency axis is defined as an offset from the excitation frequency (7.0805 GHz). Data were taken at 1.9 K in a magnetic field of 50.19 mT.}
 \label{rawdata}
\end{figure}

\begin{figure}[h]
 \includegraphics{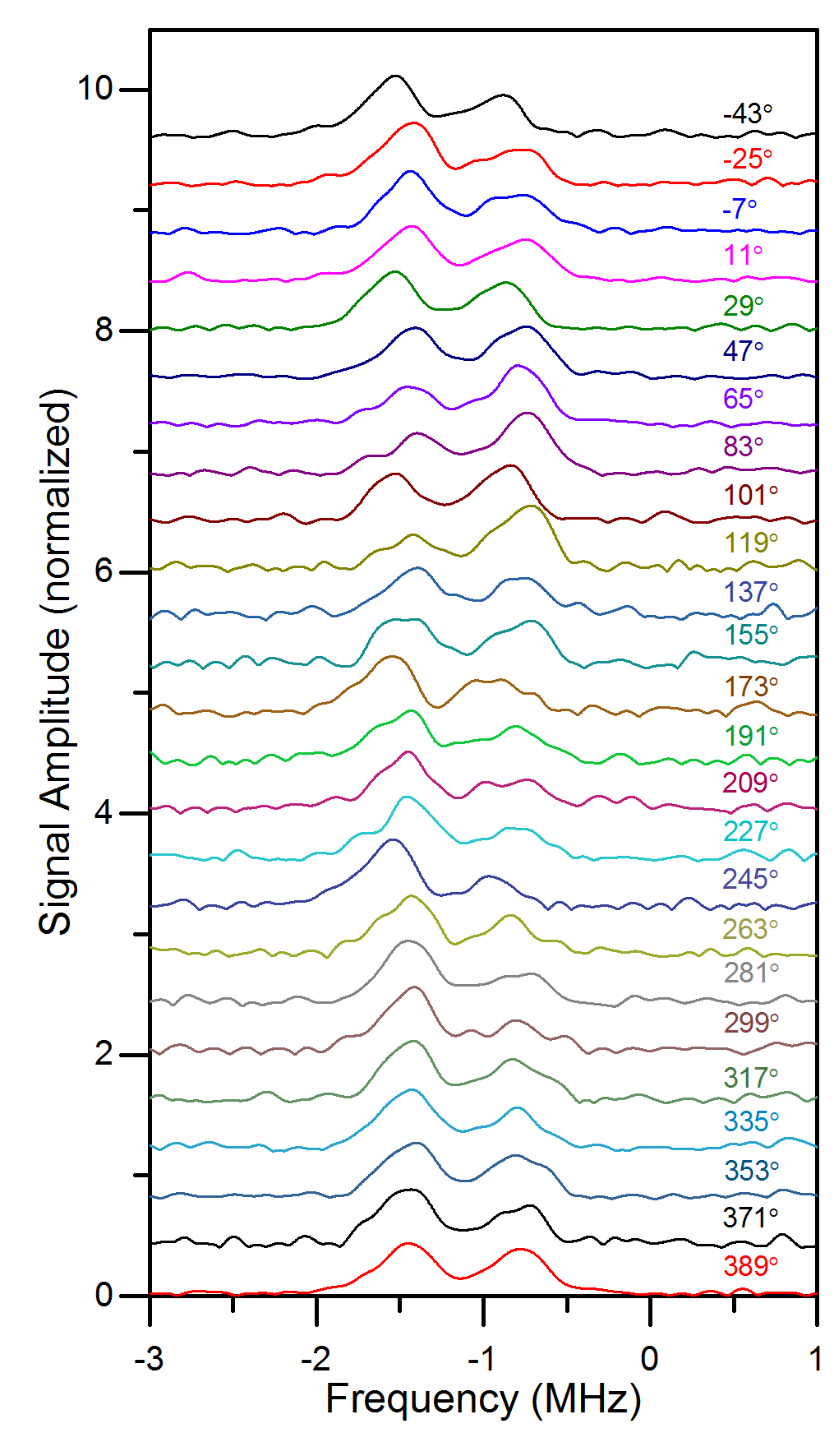}
  \caption{Fourier transformed ESR spectra as a function of relative microwave phase in the two resonators. The frequency is plotted relative to 7.0805 GHz. Data were taken at 1.9 K in a magnetic field of 50.19 mT. As the relative phase of the microwave pulses in the two resonators is varied, the relative amplitude of the allowed and forbidden transitions changes.}
 \label{waterfall}
\end{figure}

The experiments were conducted at 1.9 K in a pumped helium cryostat. As fabricated, the CPW had a resonance frequency of 7.0805 GHz (49 MHz above the clock transition) with a Q factor of 1000. The dielectric resonator's frequency was tuned to match the CPW resonance with a Q  of 7000. The peak microwave power for the two resonators was independently optimized by performing a Rabi experiment on the P donors which gave a large ESR signal due to the large number of spins. The power was then adjusted to take into account the different location of the Bi spins and the larger $B_{1}$ required near the clock transition. Regardless of which resonator was used to excite the spins, the spin echo was detected using the CPW microresonator since it is substantially more sensitive than the dielectric resonator \cite{sigillito2014, bienfaitQLS}. The spins were resonant with the CPW at $B_{0} =$ 50.19 mT ($\sim$ 30 mT away from the clock transition) and pulsed ESR was performed simultaneously on the two nearly degenerate transitions. A Carr-Purcell-Meiboom-Gill (CPMG) \cite{CPMG} sequence was employed to average multiple echoes \cite{echotrain} using the sequence ($\pi$/2-($\tau$-$\pi$-$\tau$)$\times$5) with a delay time, $\tau$, of 60 $\mu$s.The delay was chosen so that the entire pulse sequence was short relative to the onset of global magnetic field noise\cite{tyryshkin2003} and T$_2$ \cite{weis2012}. The spin echo train was signal averaged 20000 times before all 5 echoes were summed together to further improve the signal to noise ratio. An example echo is shown in Fig.\ref{rawdata}(a). Fourier transformation of the echo shape then gave an ESR signal \cite{wolfowicz2013}, as shown in Fig.\ref{rawdata}(b), with both the allowed and forbidden transitions resolved. The linewidth of both transitions is about 300 kHz and they are separated by approximately 660 kHz, consistent with simulation using Easyspin\cite{stoll2006easyspin}. The experiment was repeated 25 times while varying the phase of the microwaves in the dielectric resonator relative to the CPW. The results are plotted in Fig.\ref{waterfall}. Note that the forbidden and allowed peaks change their relative amplitude which is a signature of their selective excitation by elliptically polarized microwaves.

\begin{figure}
\includegraphics{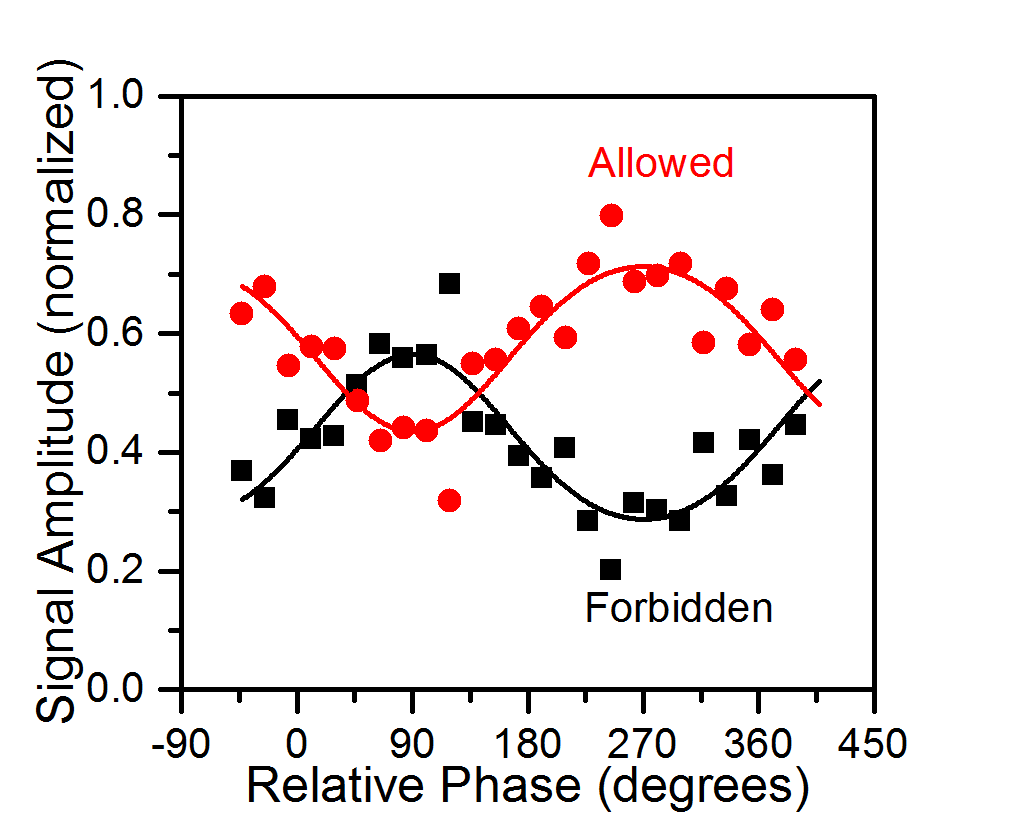}
\caption{\label{phase_dependence} The normalized amplitudes of the forbidden and allowed transitions as a function of the relative phase between microwaves in the two resonators. The solid curve represents the fit obtained using the model described in Eq. (1-3).}
\end{figure}

While we observe a phase dependence, the selectivity of our pulses is poor which is attributed to a distribution in $\vec{B}_{1,CPW}$. We modelled the system to quantitatively understand the effect of $\vec{B}_{1,CPW}$ homogeneity on the distribution of polarization and thus the phase dependence of the relative allowed and forbidden transition amplitudes. The contribution of a single spin, $i$, to the forbidden ($E_{f}^{(i)}$) and allowed ($E_{a}^{(i)}$) echo signals is given by
\begin{equation}
\label{echo}
E_{f,a}^{(i)} \propto B_{1,CPW}^{(i)} sin(\frac{g\mu_B\tau_{\pi}}{2\hbar}B^{(i)}_{1,\sigma})^ 3
\end{equation}
where $g$ is the g-factor, $\mu_B$ is the Bohr magneton, $\tau_{\pi}$ is the $\pi$ pulse length (100 ns), $\hbar$ is the reduced Planck constant, $B^{(i)}_{\sigma}$ is the amplitude of the $\sigma =$ clockwise or $\sigma =$ counterclockwise circularly polarized $\vec{B_{1}}$ components, respectively, and $B_{1,CPW}^{(i)}$ is the microwave magnetic field due to the CPW resonator at spin $i$ \cite{hoult1997}. The $B_{1,CPW}^{(i)}$ term in this expression is proportional to the spin-to-resonator coupling \cite{malissa2012}. $B^{(i)}_{1,\sigma}$ is described in terms of $\vec{B}_{1,D}$ and $\vec{B}_{1,CPW}^{(i)}$ and their relative phase difference $\phi$ as
\begin{equation}
B^{(i)}_{1,\sigma}=\frac{1}{2}\sqrt{|\vec{B}_{1,D}|^2+|\vec{B}_{1,CPW}^{(i)}|^2\pm 2|\vec{B}_{1,D}||\vec{B}_{1,CPW}^{(i)}|\cos\phi}
\label{polarizedB}
\end{equation} where $+$ corresponds to clockwise and $- $ to counterclockwise polarizations. The overall signal of each transition is simply given as a sum over each individual spin's contribution such that

\begin{equation}
\label{totalsignal}
E_{f, a} =\sum E_{f, a}^{(i)}.
\end{equation} The model was fit to the data and the resulting curves are plotted in Fig.\ref{phase_dependence} along with the experimental echo intensities. In the figure the data are normalized  so that $E_{f} + E_{a} = 1$. The simulated phase dependence agrees well with our measurements and clearly shows the polarization addressability of the forbidden and allowed transitions. 

Ideally, the data would show perfect contrast such that one transition has a signal amplitude of 1 while the other is 0. This is not the case here since the CPW inhomogeneity leads to a distribution in polarization which partially washes out the selectivity of the microwaves. It will be important for quantum devices exploiting this effect to use highly polarized microwave fields to get good addressability between qubits. Fortunately, many techniques for generating highly circularly polarized microwaves exist \cite{huchison1960, chang1964, eshbach1952} but as previously discussed, they were impractical given our sample. It is also important to note that very small samples and single donor devices\cite{Pla2012,Pla2013} will not suffer from these inhomogeneity issues since over a small volume, $\vec{B}_{1,CPW}$ is homogeneous. 

In conclusion, we have assembled a hybrid resonator consisting of a superconducting $\lambda$/4 shorted CPW and a frequency tunable dielectric resonator to controllably apply arbitrary, elliptically polarized microwaves to our spin ensemble. Using these resonators we showed that clockwise and counterclockwise circularly polarized microwaves can be used to selectively address the allowed and forbidden clock transitions for Bi donor spins in silicon. This enables the rapid manipulation of two nearly degenerate qubits in a regime where coherence times can be long, even for natural silicon. This addressability is not only important for donor-dot quantum computing schemes like the one described by Pica \textit{et al.} \cite{pica2015surface}, but also for other quantum computing architectures relying on the use of more than two donor spin states.

\begin{acknowledgments}
This work was supported by the NSF through the Materials World Network and MRSEC Programs (Grant Nos. DMR-1107606 and DMR-01420541), and the ARO (Grant No. W911NF-13-1-0179).
\end{acknowledgments}

\providecommand{\noopsort}[1]{}\providecommand{\singleletter}[1]{#1}%

\end{document}